\documentstyle[12pt]{article}
\newlength{\mathspace}

\def\np#1{ Nucl. Phys. {\bf  B#1} }

\def\pl#1{ Phys. Lett.  {\bf B#1}}
\def\cmp   { Commun. Math. Phys. }

\def\mpl   { Mod. Phys. Lett. }
\def\del {\partial}
\def\chat {{\hat c}}

\def\begineq{\begin{equation}}
\def\endeq{\end{equation}}
\def\eqabegin{\begin{eqnarray}}
\def\eqaend{\end{eqnarray}}
\def\nn{\nonumber}

\def\parbigskip        {  \par\bigskip  }
\def\parmedskip        {  \par\medskip  }
\def\parsmallskip      {  \par\smallskip  }
\def\parbigskipn        {  \par\bigskip\noindent  }

\begin{document}
\baselineskip=0.7cm
\setlength{\mathspace}{2.5mm}



\begin{titlepage}

    \begin{normalsize}
     \begin{flushright}
                 UT-Komaba/94-14 \\
                 hep-th/9406171 \\
                 June 1994
     \end{flushright}
    \end{normalsize}
    \begin{LARGE}
       \vspace{1cm}
       \begin{center}
         Topological Structure in \\
         ${\hat c} = 1$ Fermionic String Theory\\
       \end{center}
    \end{LARGE}

  \vspace{5mm}

\begin{center}
           Shinji H{\sc irano}
           \footnote{E-mail address:
              hirano@hep1.c.u-tokyo.ac.jp}
           \footnote{JSPS Research Fellow}
                \ \  and \ \
           Hiroshi I{\sc shikawa}
           \footnote{E-mail address:
              ishikawa@hep1.c.u-tokyo.ac.jp}
           \footnote{JSPS Research Fellow}\\
      \vspace{4mm}
        {\it Institute of Physics, College of Arts and Sciences} \\
        {\it University of Tokyo, Komaba}\\
        {\it Meguro-ku, Tokyo 153, Japan}\\
      \vspace{1cm}

    \begin{large} ABSTRACT \end{large}
        \par
\end{center}
\begin{quote}
 \begin{normalsize}
\ \ \ \
$\chat=1$ fermionic string theory, which is considered as a
fermionic string theory in two dimension, is shown to
decompose into two mutually independent parts, one of which can be
viewed as a topological model and the other is irrelevant for the
theory. The physical contents of the theory is largely governed by this
topological structure,
and the discrete physical spectrum of
$\chat=1$ string theory is naturally explained as the physical spectrum
of the topological model. This topological structure turns out to be
related with a novel hidden
$N=2$ superconformal algebra (SCA) in the enveloping algebra of the
$N=3$ SCA in fermionic string theories.
 \end{normalsize}
\end{quote}

\end{titlepage}
\vfil\eject



The remarkable feature of the two-dimensional string theory is that
only the tachyon field is left as a continuous physical
degree of freedom, while most of the excitation modes of the string
are eliminated and there remains an infinite set of discrete physical
states as their remnants \cite{DSC,LZ,BMP1,Wit,IO,BMP2}.
The drastic reduction of physical degrees of freedom makes the theory
so simple that the two-dimensional string theory may provide a good
testing ground for discussing the non-perturbative formulation and the
dynamical principle of string theory.

The reduction of physical degrees of freedom suggests that the
two-dimensional string theory can be described as a topological model,
which may give us an alternative formulation of string theory.
Indeed, several authors showed the correspondence of $c=1$ string
theory, which is considered as a bosonic string theory in
two-dimensional space-time, to certain topological field theories
\cite{top1,top2,top3,top4,top5,top6}. However they have not been able
to clarify the role of the discrete states which we consider to be
deeply connected with the topological nature of two-dimensional string
theories.

In the recent work \cite{IK}, a relation between $c=1$
string theory and a topological sigma model was suggested, in which
the former was shown to be a kind of realization, or `bosonization',
of the latter.
In this formulation, the existence of the discrete states was crucial.
Namely, the target space coordinate and the
gauge ghost of the topological sigma model were realized by one of
the ground ring generators \cite{Wit} and a discrete tachyon,
respectively. Consequently, all the discrete states in $c=1$
string theory were reproduced as the physical states of the
topological sigma model.

In this letter, we apply the analysis performed on $c=1$ string theory
\cite{IK} to the case of $\chat=1$ fermionic string theory, which is
considered as a fermionic string theory in two dimension. We
have found that $\chat=1$ string theory contains a subspace which can
be viewed as a topological model. This topological model is essentially
the same one as obtained in the $c=1$ case. The physical spectrum of
the topological model turns into the discrete physical spectrum in
$\chat=1$ string theory. Therefore, one can say that $\chat=1$ string
theory is largely governed by the topological model as for the physical
contents of the theory. As we will show, the
$N=3$ superconformal symmetry in fermionic string theories \cite{BLNW}
plays an important role in this topological structure. Namely, we
identify the (twisted)
$N=2$ superconformal algebra (SCA) of the topological model with a hidden
$N=2$ SCA in the enveloping algebra of the $N=3$ SCA. Associated with
this fact, we found a novel automorphism of $N=3$ SCA in the enveloping
algebra.

A notion that $\chat=1$ string theory can be regarded as a
bosonization of a topological model was previously pointed out also by
Distler \cite{Distler}. The difference between his and our approaches
shows up in the treatment of the diffeomorphism ghosts, i.e., upon
bosonization, we combine the ghosts with the other fields into the
form of the discrete states in $\chat=1$ string theory, while his
approach leaves the ghosts unchanged. Therefore, the appearance of the
discrete physical spectrum is more transparent from our point of
view.

%

%

\parmedskip
Now we begin our analysis by giving some notations and conventions.
$\chat=1$ fermionic string theory consists of two free scalars, the
matter field $\phi^M$ and the Liouville field $\phi^L$, their
superpartners $\psi^M$ and $\psi^L$, the gauge fixing ghosts $b$ and
$c$ for the diffeomorphism invariance, and their superpartners $\beta$
and $\gamma$\footnote[2]{Only the holomorphic part will be considered
throughout this letter.}.
The energy-momentum tensor $T$ and the supercurrent
$G$ of $\chat=1$ string theory are given by
\begin{eqnarray}
\label{T}
  T &=& T^M + T^L + T^G \, ,\\
  G &=& G^M + G^L + G^G \, ,
\end{eqnarray}
where the superscripts $M$, $L$ and $G$ denote the matter, the Liouville
and the ghost sector respectively, while $T^i$ and $G^i$ ($i=M, L, G$)
stand for the energy-momentum tensor and the supercurrent of each
sector.
In the above equation
\begin{eqnarray}
  T^i &=& -{1 \over 2} (\del \phi^i)^2 + i \lambda^i \del^2 \phi^i -
{1 \over 2} \psi^i \del \psi^i \, ,\,(i=\mbox{M,L})\, , \\
  T^G &=& c\del b + 2\del cb - {1 \over 2} \gamma
\del \beta - {3 \over 2} \del \gamma\beta \, , \\
  G^i &=& i \del \phi^i \psi^i + 2\lambda^i \del \psi^i \,
    ,\,(i=\mbox{M,L})\, , \\
  G^G &=& b\gamma - 3 \del c\beta - 2 c\del\beta  \, ,
\end{eqnarray}
where $\lambda^M=0$ and $\lambda^L=-i$.

The BRST charge $Q_{\chat=1}$ of
$\chat=1$ string theory is expressed as
\begin{equation}
\label{BRST}
Q_{\chat=1} = \oint\! dz\, j_{BRST} =
  \oint \! dz \left[ c\left( T^M +
  T^L + {1 \over 2}T^G\right) -{1 \over 2}\gamma\left( G^M + G^L +{1
  \over 2}G^G\right)\right].
\end{equation}
As usual, these fields have to be bosonized in order to treat
different pictures \cite{FMS}. We give the explicit form of the
bosonization for $\chat=1$ string theory as follows: %
\eqabegin
\label{bosonization}
& &\psi^{\pm}=e^{\pm ih}\,\mbox{\bf I}, \nn\\
& &\gamma=\eta e^{\phi}\,\mbox{\bf I},\qquad\quad\beta=\del\xi
e^{-\phi}\,\mbox{\bf I}, \\
& &h(z)h(w)\;\; \sim\;\; -ln(z-w) \;\;\sim \;\;\phi(z)\phi(w).\nn
\eqaend
In the above equation, $\psi^{\pm}={1 \over \!\!\sqrt{2}}\left(\psi^M\pm
i\psi^L\right)$, and $h$ and $\phi$ are free scalars. ($\xi,\eta$)
denotes a pair of the fermionic first-order fields with spin-(0,1).
{\bf I} is a cocycle factor which anti-commutes with $bc$-ghosts and
commutes with the other fields. Then one can easily check that the
bosonized form (\ref{bosonization}) of the fundamental fields in
$\chat=1$ string theory have the correct mutual statistics.

The spin field $\Sigma$ that creates the vacuum for the canonical
Ramond sector from the SL(2,$\bf{C}$) invariant vacuum is expressed as
$\Sigma=e^{-{ih\over 2}}e^{-{\phi \over 2}}\,\mbox{\bf I$_R$}$.
Here, the cocycle factor {\bf I$_R$} is determined so that the spin
field has the definite statistics with respect to the BRST charge
(\ref{BRST}). The explicit form of {\bf I$_R$} is given
by
\eqabegin
\mbox{\bf I$_R$}=e^{{\pi i \over 2}(N_h-N_{\phi})},
\eqaend
where $N_h={1 \over 2\pi i}\oint\! dz\, i\del h$, and
$N_{\phi}=-{1\over 2\pi i}\oint\! dz\,\del\phi$.
With this choice for $\mbox{\bf I$_R$}$, the statistics of $\Sigma$
is bosonic with respect to the BRST charge.
%
%
\parmedskip
We turn to the construction of a topological structure in $\chat=1$
fermionic string theory.
Our starting point is a topological model
consisting of the fields $B$, $C$, $x$ and ${\bar p}$, which is
essentially a (twisted) supersymmetric first-order system. These fields
satisfy the following operator product expansion (OPE)
\begin{equation}
  \label{fundamentalOPE}
  \begin{array}{rcccl}
  B(z) C(w)  & \sim & {\displaystyle \frac{1}{z-w}}  & \sim &
  C(z) B(w) \, , \\[\mathspace]
  \bar{p}(z) x(w) & \sim & {\displaystyle \frac{1}{z-w}} & \sim &
  - x(z) \bar{p}(w) \, .
  \end{array}
\end{equation}
The other combinations vanish.
The generators of the twisted $N=2$ SCA which characterize this
topological model are expressed as follows:
\begin{equation}
  \label{topN=2}
  \begin{array}{rcl}
  T_{top} &=& \del x\, \bar{p} - B \del C\, ,\\[\mathspace]
  G^+_{top} &=& -\del C\, x\, ,\\[\mathspace]
  G^-_{top} &=& B \bar{p}\, ,\\[\mathspace]
  J_{top} &=& \bar{p} x.
  \end{array}
\end{equation}
The BRST current of this model is nothing but the
supercurrent $G^+_{top}$.
Hence the BRST charge $Q_{top}$ of the model is
expressed as
\begin{equation}
  Q_{top} = \oint\! dz (-\del C\, x) \label{topBRST} \, .\\
\end{equation}

This topological model is previously shown \cite{IK} to yield
$c=1$ bosonic string theory by appropriately identifying the fields
with those in the $c=1$ Fock space \cite{IK}.
In \cite{IK}, $x$ and $C$ were realized by the discrete operators of
$c=1$ string theory, and the discrete physical spectrum of $c=1$ string
theory was derived from the physical spectrum of the topological model.
One can say that the existence of the discrete physical spectrum is
intimately related to the existence of the corresponding topological
model. We can expect that
the similar program is also possible to carry out in the present case,
since the structure of the discrete states is almost the same as that
in the $c=1$ case. As we will show below, this is the case.

First, we introduce a set of four fields,
the explicit form of which is given by
\begin{eqnarray}
    B_{st} &=&
   \left(b\Sigma^{(1/2,1/2)}-\sqrt{2}\del\xi\Sigma^{(-1/2,-1/2)}\right)
   e^{\; iX^+ \over\!\!\sqrt{2}}\mbox{\bf I$_R$}^{-1}, \label{Bst} \\
  C_{st} &=&
   c\Sigma^{(-1/2,-1/2)}e^{-{ \; iX^+\over \!\!\sqrt{2}}}\,
   \mbox{\bf I$_R$}, \label{C} \\ \label{Cst}
  x_{st} &=& \left[\frac{}{}\!\! \left(\sqrt{2}i\del X^-
  -bc\right)\Sigma^{(1/2,1/2)}-\sqrt{2}c\del\xi\Sigma^{(-1/2,-1/2)}
  \right.\nn\\
  & &\qquad\qquad\qquad\left. +{1 \over\!\!\sqrt{2}}
  b\eta\Sigma^{(-1/2,3/2)}+\Sigma^{(-3/2,1/2)}\right] e^{ \; iX^+
  \over\!\!\sqrt{2}}\,\mbox{\bf I$_R$}^{-1} \,\\ \label{xst}
  {\bar p}_{st} &=&  {1\over2}
   \left(\Sigma^{(-1/2,-1/2)}-\sqrt{2}c\del\xi\Sigma^{(1/2,-3/2)}\right)
   e^{ -{\; iX^+ \over\!\!\sqrt{2}}}\mbox{\bf I$_R$} \, . \label{pst}
\end{eqnarray}
The fields $C_{st}$ and $x_{st}$ are physical with respect to the
BRST charge (\ref{BRST}) of $\chat=1$ string theory. $C_{st}$ is one
of the discrete tachyon while $x_{st}$ is the ground ring
generator \cite{BMP2} (The subscript `$st$' means that these fields are
defined in the Fock space of $\chat=1$ string theory).
It is clear from the appearance of the cocycle factor $\mbox{\bf
I$_R$}$ that the above fields are in the Ramond sector
(Note that the
ground ring generators appear in the Ramond sector for $\chat\le 1$
fermionic string theory \cite{BMP2}).
For fermionic string theories, it is well-known that there exists the
picture-changing isomorphism \cite{FMS} of the physical operators. We
have chosen the picture of $x_{st}$ and $B_{st}$ as $+1/2$ and that of
$C_{st}$ and ${\bar p}_{st}$ as
$-1/2$\footnote[2]{We assign picture $-1/2$ to the canonical Ramond
sector and $0$ to the $SL(2,{\bf{C}})$ vacuum.}.

These fields realize, in the $\chat=1$ Fock space, the fields
(\ref{fundamentalOPE}) of the topological model.  In
fact, one can show that these fields satisfy the following OPE
\begin{equation}
  \begin{array}{rcccl}
  B_{st}(z) C_{st}(w)  & \sim & {\displaystyle \frac{1}{z-w}} & \sim &
  C_{st}(z) B_{st}(w) \, , \\[\mathspace]
  \bar{p}_{st}(z) x_{st}(w) & \sim & {\displaystyle \frac{1}{z-w}}
  & \sim & - x_{st}(z) \bar{p}_{st}(w) \, ,
  \end{array}
\end{equation}
which is exactly the same form as in eq.(\ref{fundamentalOPE}).
If we identify these fields with the fields (\ref{fundamentalOPE}) of
the topological model,
we can also realize the twisted $N=2$ SCA
(\ref{topN=2}) in the $\chat=1$ Fock space.
In the case of $c=1$ bosonic string theory \cite{IK}, this procedure
yielded a twisted $N=2$ SCA in the $c=1$ Fock space with the BRST
current and the total energy-momentum tensor as the generators $G^+$
and $T$ respectively.
More specifically,
by substituting an appropriate expression for the fields of the
topological model into eq.(\ref{topN=2}),
the energy-momentum tensor $T_{top}$ in (\ref{topN=2})
of the topological model turned into the total energy-momentum tensor
of $c=1$ string theory and the supercurrent, or the BRST current,
$G^+_{top}$ in (\ref{topN=2}) into the BRST current of string theory.
The fact that the energy-momentum tensor of the topological model
yields the one of $c=1$ string theory indicates that the topological
model is realized not in a part of but in the entire Fock space of the
string theory. On the other hand, the correspondence of the BRST
currents implies the equivalence of the physical spectra of both
models. These two facts enabled us to regard
$c=1$ string theory as a realization of the topological
model (\ref{fundamentalOPE}).
We can expect that this topological model describes $\chat=1$ string
theory as for, in particular, its BRST charge and physical spectrum, in
the same manner.
However, the direct
application of this procedure fails;
$Q_{\chat=1} \ne \oint (-\del C_{st}\, x_{st})$.

The situation can be remedied by replacing the usual derivative $\del$
with a `modified' one ${\cal L}_{-1}$, namely we can obtain the
following expression
\begin{equation}
 \label{top2string}
 Q_{\chat=1} = \oint\! dz (-{\cal L}_{-1}C_{st}\, x_{st}) \, .
\end{equation}
In the above equation, ${\cal L}_{-1}C_{st}\, x_{st}$ is the finite part
of the OPE of ${\cal L}_{-1}C_{st}$ with $x_{st}$.
The operator ${\cal L}_{-1}$ is defined as $(-1)$-mode of a
`modified' energy-momentum tensor ${\cal L}(z)$ which is given by
\begin{equation}
  {\cal L}(z) = \sum_n {\cal L}_{-n} z^{n-2} =
  \{Q_{\chat=1}, b^\star (z) \} \, ,
  \label{defL}
\end{equation}
where $b^\star$ takes the following form
\eqabegin
 \label{eqn:bstar}
b^{\star} &=& {1 \over 2}\left( b +
\sqrt{2}\beta\left( \psi^+ -\psi^-\right) + 2c\beta^2\right)\\
&=&{1\over 2}\left( b + \sqrt{2}\del\xi\left(
e^{ih-\phi}-e^{-ih-\phi}\right)+2c\del^2\xi\del\xi
e^{-2\phi}\right)\, ,\nn
\eqaend
and has a vanishing OPE with itself.
The above equation (\ref{defL}) is considered
as a modified version of the well-known relation
\begin{equation}
  T(z) = \{Q_{\chat=1}, b(z)\} \, .
\end{equation}
Actually, by simple but tedious calculation, one can show that the
field ${\cal L}(z)$ satisfies the Virasoro
algebra with vanishing central charge; a necessary property for
an energy-momentum tensor of a topological model. This is the reason
we call the field in eq.(\ref{defL}) as a modified energy-momentum
tensor. Accordingly the $(-1)$-mode of ${\cal L}$ can be considered as
a modified derivative. As is shown in eq.(\ref{top2string}), it is
necessary to replace the ordinary derivative $\del$ with the modified
one ${\cal L}_{-1}$ in order to account for the physical contents of
$\chat=1$ string theory in terms of the topological model.

This energy-momentum tensor ${\cal L}$ describes a part of degrees of
freedom in $\chat=1$ string theory. In order to see this, let us
decompose the total energy-momentum tensor (\ref{T}) of string theory
as follows:
\begin{equation}
  \label{decomposition}
  T = {\cal L} + \tilde{\cal L} \, .
\end{equation}
Since $b^\star$ is primary with respect to the total energy-momentum
tensor $T$, it follows that, from eq.(\ref{defL}), ${\cal L}$ is
primary with respect to $T$ (Note that $T$ commutes with the BRST
charge $Q_{\chat=1}$). From this and the fact that ${\cal L}$
satisfies the Virasoro algebra, one can show that $\tilde{\cal L}$
also satisfies the Virasoro algebra with vanishing central charge and
commutes with ${\cal L}$:
\begin{eqnarray}
  {\cal L}(z) {\cal L}(w) &\sim &
   \frac{2 {\cal L}(w)}{(z-w)^2} + \frac{\del {\cal L}(w)}{z-w}\, ,\\
  \tilde{\cal L}(z) \tilde{\cal L}(w) &\sim &
   \frac{2 \tilde{\cal L}(w)}{(z-w)^2} +
   \frac{\del \tilde{\cal L}(w)}{z-w}\, ,\\
  {\cal L}(z) \tilde{\cal L}(w) &\sim & 0 \, .
\end{eqnarray}
Hence, the field $\tilde{\cal L}$ can also be regarded as an
energy-momentum tensor independent of ${\cal L}$.
The decomposition (\ref{decomposition}) of the total energy-momentum
tensor implies that the $\chat=1$ Fock space ${\cal F}_{\chat=1}$
can be viewed as a direct
product of two Fock spaces,
which we denote as ${\cal F}$ and $\tilde{\cal F}$,
corresponding to two mutually independent
energy-momentum tensors ${\cal L}$ and $\tilde{\cal L}$:
\begin{equation}
  \label{Fock_decomposition}
  {\cal F}_{\chat=1} = {\cal F} \bigotimes \tilde{\cal F} \, .
\end{equation}
Since ${\cal L}$ and $\tilde{\cal L}$ have vanishing central charge,
both of these two sectors are considered to be topological.

The fact that the modified derivative ${\cal L}_{-1}$ appears in
eq.(\ref{top2string}) instead of the ordinary one $\del$ suggests that
the energy-momentum tensor
we should use in order to keep the structure of the topological model
is not the usual but the modified one (\ref{defL}).
In other words, the topological model (\ref{fundamentalOPE}) is
realized in the subspace ${\cal F}$, which is characterized by
${\cal L}$, of the $\chat=1$ Fock space.
In fact, by identifying the derivative in the
topological model with this modified one ${\cal L}_{-1}$, the $N=2$
SCA (\ref{topN=2}) in the topological model yields a twisted $N=2$
SCA with ${\cal L}$ as
the energy-momentum tensor:
\begin{equation}
  \label{stringN=2}
  \begin{array}{rc@{\hspace{2mm}}c@{\hspace{2mm}}cl}
  T_{st} &=& {\cal L}_{-1} x_{st}\: \bar{p}_{st}
    - B_{st}\: {\cal L}_{-1} C_{st} &=& {\cal L}\, ,\\[\mathspace]
  G^+_{st} &=& -{\cal L}_{-1} C_{st}\: x_{st} &=&
    {\displaystyle
      j_{BRST} - \del\left(
      {1 \over 2}\del c + c\del\phi^L+{i \over 2} \gamma\psi^L
      -{1 \over 4}c\beta\gamma\right)} ,\\[\mathspace]
  G^-_{st} &=& B_{st} \bar{p}_{st} &=& b^\star \, ,\\[\mathspace]
  J_{st} &=& \bar{p}_{st} x_{st} &=& -b c - \beta \gamma + \del \phi^L
    \, .
  \end{array}
\end{equation}

Thus we have obtained a topological structure in $\chat=1$ string
theory which can be regarded as a realization of the topological model
(\ref{fundamentalOPE}). Although this topological structure is
realized in the subspace ${\cal F}$ rather than the entire Fock space
of string theory, the physical contents of string theory is largely
governed by this structure.
The physical spectrum of the topological model (\ref{fundamentalOPE})
consists of zero modes of $x$ and $C$.
Through the realization (\ref{Bst})-(\ref{pst}) of the topological
model, this physical spectrum gives rise to that of the topological
model realized in ${\cal F}$ (\ref{stringN=2}), which is expressed by the
zero modes of $x_{st}$ and $C_{st}$. However, these are
nothing but the discrete physical operators, and are physical not only in
${\cal F}$ but also in the entire $\chat=1$ Fock space.
Hence, the discrete physical spectrum of $\chat=1$ string theory is
explained in terms of the topological model through the topological
structure in ${\cal F}$ (\ref{stringN=2}). One can say that, at least
for the discrete spectrum, the subspace $\tilde{\cal F}$ in the
decomposition (\ref{Fock_decomposition}) is irrelevant, and that the
physical contents of the theory is governed by the topological
structure in ${\cal F}$ (\ref{stringN=2}) or the topological model
(\ref{fundamentalOPE}) itself.

\parbigskip
%

Although the geometrical meaning of the decomposition
of the Fock space (\ref{Fock_decomposition}), or the energy-momentum
tensor (\ref{decomposition}), has
not yet been clarified, we do have the algebraic
meaning of ${\cal L}(z)$, i.e. it can be characterized by a twisted
$N=3$ SCA in $\chat=1$ string theory.

As was discussed in \cite{BLNW}, any fermionic string theory has a
twisted $N=3$ superconformal symmetry. In the following, we will
give an explicit representation of the twisted $N=3$
SCA in $\chat=1$ fermionic string theory.  The
three supercurrents ($G^+,G^3,G^-$) form a $SU(2)$ triplet which
consists of the BRST current with a suitable total derivative
modification, $N=1$ supercurrent, and the anti-ghost $b$,
respectively. Namely,
\eqabegin
G^+&=&\sqrt{2}\left[ c\left( T^M + T^L + {1 \over
2}T^G\right) -{1 \over 2}\gamma\left( G^M + G^L
+{1 \over 2}G^G\right)\right.\nn\\ \qquad\qquad\qquad
& &+\del\left\{ {1 \over 2}\del c+{1 \over
4}c\beta\gamma -q_1\left(
2ci\del\phi^M-\psi^M\gamma\right)\right.\nn\\
\qquad\qquad\qquad& &\left.\left.-q_2\left(
2i\del c+2ci\del\phi^L-\psi^L\gamma\right)
\right\}\right],\label{eqn:n3al1}\\
G^3&=&G^M + G^L + G^G,\\
G^-&=&\sqrt{2}b.
\eqaend
Here and hereafter $q_1=\pm(iq_2-{1 \over 2})$.

The energy-momentum tensor of this algebra is that of
$\chat=1$ fermionic string theory (\ref{T}).
The $SU(2)$ currents ($J^+,J^3,J^-$) are identified with the
$G^3$-super partner of the modified BRST current $G^+$, the
modified ghost number current, and the $G^3$-super partner of
$G^-$, respectively:
\eqabegin
J^+&=&-c\left(
G^M+G^L +b\gamma-\del c\beta\right)
+{1 \over 2}\beta\gamma^2+\del\gamma\nn\\
\qquad\qquad\qquad
& &\!\!\!-2q_1\left( \gamma i\del\phi^M-\del
c\psi^M-2c\del\psi^M\right) \nn\\
\qquad\qquad\qquad
& &\!\!\!+2q_2\left( \del c\psi^L+2c\del\psi^L
-\gamma i\del\phi^L-2i\del\gamma\right)\!\! ,\\
J^3&=&-bc-\beta\gamma+2q_1i\del\phi^M
+2q_2 i\del\phi^L,\\
J^-&=&-2\beta.
\eqaend
There is also an additional fermion which is the $G^3$-super
partner of the $U(1)$ current $J^3$:
\eqabegin
\Psi=-\beta c+2q_1\psi^M+2q_2\psi^L.
\label{eqn:n3al2}
\eqaend
The central charge of this $N=3$ SCA is equal to $c^{N=3}=3-12iq_2$.
%
%
In the following, we shall show that the twisted $N=2$ SCA in
(\ref{stringN=2}) is nothing but a hidden $N=2$ algebra
in the enveloping algebra of this
twisted $N=3$ SCA.

First, note that the supercurrent $G^+_{st}$
and the $U(1)$ current $J_{st}$ of the topological structure
(\ref{stringN=2})
are equal
to those of the twisted $N=3$ SCA given above, the parameters $q_1$
and $q_2$ of which are set to be $0$ and $-i/2$ respectively:
\eqabegin
  \label{G+N=3}
  \sqrt{2} G^+_{st}(z) &=& G^+(z),\\
  \label{JN=3}
  J_{st}(z) &=& J^3(z).
\eqaend
Here and hereafter the generators without subscripts are
those of the twisted $N=3$ SCA.
Next let us examine the modified anti-ghost $b^{\star}=G^-_{st}$
in the topological structure. One can easily find that this can be
written in terms of the generators of the twisted $N=3$ SCA,
namely,
\begin{equation}
  \label{G-N=3}
  \sqrt{2} G^-_{st}(z)={1 \over 2}\left( G^-(z) +\sqrt{2}\Psi
  J^-(z)\right).
\end{equation}
Finally, we can read off the expression of the
energy-momentum tensor ${\cal L}(z)$ of the topological structure in
terms of the $N=3$ SCA from the single pole of the OPE
$G^+_{st}(z)G^-_{st}(w)$:
\begin{equation}
  \label{LN=3}
  {\cal L}(z)={1 \over 2}\left( T(z) + {1 \over 2}J^-J^+(z) +
  G^3\Psi(z)\right).
\end{equation}
Thus, the $N=2$ SCA (\ref{stringN=2}) in the subspace ${\cal F}$
has a simple expression in terms of the
twisted $N=3$ SCA in $\chat=1$ fermionic string
theory.

The above expressions (\ref{G+N=3})-(\ref{LN=3}) suggest the existence
of an $N=2$ SCA in the enveloping algebra of a generic $N=3$ SCA.
In fact, apart from $\chat=1$ string
theory, we can show that in general there exists a hidden $N=2$ SCA
in the enveloping algebra of an $N=3$ SCA,
the generators of which read
\eqabegin
\tilde{T} &=&
                        {1 \over 1+{\alpha \over \sqrt{2}}}
      \left[T +{\alpha \over 2\sqrt{2}}\left(J^-J^+ +2G^3
      \Psi\right)\right]\, , \\
\tilde{G}^+ &=& G^+, \\
\tilde{G}^- &=& {1 \over 1+{\alpha \over
  \sqrt{2}}}\left( G^- + \alpha\Psi J^-\right), \\
\tilde{J} &=& J,
\eqaend
where $\alpha=-{3\sqrt{2} \over c}$ and the central charge of this
algebra equals to $c$. In the above equation, all the generators
are expressed in the twisted form for both a hidden $N=2$ SCA and
the $N=3$ SCA. One can easily verify that the case of
$\chat=1$ string theory (\ref{G+N=3})-(\ref{LN=3}) corresponds to
$\alpha=\sqrt{2}$; the $N=2$ SCA in (\ref{stringN=2}), which
realizes the topological model in $\chat=1$ string theory, is the
hidden $N=2$ SCA in the enveloping algebra of the $N=3$ SCA.

Since an $N=3$ superconformal algebra
has a symmetry under the replacements $T\to T$, $G^{\pm}\to G^{\mp}$,
$G^3\to -G^3$, $J^3\to -J^3$, $J^{\pm}\to J^{\mp}$, and $\Psi\to
\Psi$, we can also obtain another $N=2$ SCA with
these replacements.
Associated with this fact,
we can obtain an
automorphism in the enveloping algebra of an $N=3$ SCA:
\eqabegin
\begin{array}[b]{lll}
\mbox{Energy-momentum tensor:}& \tilde{T}(z)=T(z),& \\
\mbox{Supercurrents:}&\tilde{G}^{\pm}(z)=G^{\pm}(z)+\alpha\Psi
J^{\pm}(z),& \tilde{G}^3(z)=G^3(z)+\sqrt{2}\alpha\Psi J^3(z),\\
\mbox{$SU(2)$ currents:}&\tilde{J}^{\pm}(z)=J^{\pm}(z),&
\tilde{J}^3(z)=J^3(z),\\
\mbox{Fermion:}&\tilde{\Psi}(z)=-\Psi(z),&\nn
\end{array}
\eqaend
where $\alpha=-{3\sqrt{2} \over c}$ and the central charge of this algebra
equals to $c$, and all the generators are of untwisted form.

\parbigskip
To summarize, we have shown that $\chat=1$ string theory consists of
two parts ${\cal F}$ and $\tilde{\cal F}$, one of which (${\cal F}$)
can be regarded as a realization of a topological model. The physical
spectrum of the topological model turns into the discrete physical
spectrum of
$\chat=1$ string theory in this realization. Hence, we can say that the
other part $\tilde{\cal F}$ is irrelevant and $\chat=1$ string theory is
largely governed by the topological model as for the physical contents
of the theory. We have also found that this topological structure in
the string theory can be understood as a hidden $N=2$ SCA in the
enveloping algebra of the $N=3$ SCA in the fermionic string theory.

In \cite{IK}, $c=1$ string theory is shown to be related with a
topological model. Since $\chat=1$ string theory has a richer symmetry
than $c=1$ string theory, one may naively consider that the topological
model corresponding to $\chat=1$ string theory possesses an extended
symmetry compared with the $c=1$ case.
In contrast with this expectation, we have shown that a topological
model for the $\chat=1$ case is essentially the same one as obtained
in the $c=1$ case. The effect of the additional symmetry appears as
the modification of the energy-momentum tensor, or the derivative.
Indeed, this modification is
possible because of the existence of the $N=3$ SCA, which results from
the world-sheet supersymmetry in $\chat=1$ string theory.

Although we have obtained the algebraic meaning of the modified
energy-momentum tensor ${\cal L}$, it is still unclear what this
modification implies in the context of string theory and how the
topological model relates with $\chat=1$ string theory. In order to
answer to these questions, we need understand
the role of the subspace
$\tilde{\cal F}$. This sector is independent of the relevant one
${\cal F}$ and seems not to contribute to the physical contents of the
model. Moreover, the energy-momentum tensor $\tilde{\cal L}$ for this
sector has vanishing central charge. This situation reminds us of the
$G/H$ coset model \cite{GKO,KS}.
The $G/H$ coset model consists of a $G$-current algebra, the gauge
field and the gauge ghosts. The entire Fock space is split into two
parts; one is the $G/H$ part and the other corresponds to the
$H$-current algebra, the gauge fields and the gauge ghosts. The latter
is topological and gauged away by making use of the BRST charge for
gauge fixing. This structure is similar to the present case if we regard
${\cal F}$ as the $G/H$ sector and $\tilde{\cal F}$ as the topological
sector which is gauged away.
If this is the case,
the topological structure we have found is a string theory with a
certain gauge symmetry, and the fact that the $\tilde{\cal F}$ is
irrelevant for the physical contents is naturally explained from the
point of view of gauge fixing.
In order to establish this point of view for the topological structure
in $\chat=1$ string theory, we need more careful analysis for the
irrelevant sector $\tilde{\cal F}$, in particular identification of
the BRST charge for the gauge fixing.

Another possible interpretation of the irrelevant sector
$\tilde{\cal F}$ is concerned with the embedding of bosonic string
theories into a fermionic string theory
\footnote{We thank to M. Kato for drawing our attention to this
point.}.  As was mentioned above, the
topological structure (\ref{stringN=2}) we have observed in $\chat=1$
string theory is essentially the same one as obtained in the $c=1$ case
\cite{IK}. In this sense, this topological structure can be viewed as
an embedding of $c=1$ bosonic string theory into $\chat=1$ fermionic
string theory.
Recently, it has been shown
\cite{BV,IK2} that all the bosonic string theories are considered to be
equivalent with a particular class of fermionic string theories. More
precisely, the Fock space of this fermionic string theory is possible
\cite{IK2} to decompose into the tensor product of that of the bosonic
string theory and a topological sector which is irrelevant for the
physical contents of the theory. In other words, bosonic string theories
can be embedded into fermionic string theories. Since $c=1$ string
theory is one of the bosonic string theories, we can apply this
procedure to $c=1$ string theory and embed it into a fermionic string
theory. Thus we have two embeddings of $c=1$ string theory into
a fermionic string theory; one is our topological structure
(\ref{Fock_decomposition}) and the other is based on the method in
\cite{BV,IK2}. It is highly plausible that these two embeddings are
actually identical and that $\chat=1$ string theory belongs to the
particular class of fermionic string theories obtained in \cite{BV}.
Although these two kinds of fermionic string theories are apparently
distinct from each other, it may be possible to relate them via, say,
a similarity transformation as utilized in \cite{IK2}.

\parbigskipn\parsmallskip
We would like to thank M. Kato for many useful discussions and
comments on the manuscript.
We are indebted to A. Fujitsu for his package
{\tt ope.math} \cite{OPE}. This work is partly supported by
the Japan Society for the Promotion of Science.

\end{document}